\documentclass[twocolumn,aps,prl,superscriptaddress]{revtex4-2}
\usepackage{amsmath}
\usepackage{graphicx}
\usepackage{MnSymbol}
\usepackage{hyperref}
\usepackage{fancyhdr}
\usepackage{booktabs}
\usepackage{multirow}
\usepackage{parskip}

\begin{document}

\title{Brownian friction dynamics: fluctuations in sliding distance}

\author{R. Xu}
\affiliation{State Key Laboratory of Solid Lubrication, Lanzhou Institute of Chemical Physics, Chinese Academy of Sciences, 730000 Lanzhou, China}
\affiliation{Peter Gr\"unberg Institute (PGI-1), Forschungszentrum J\"ulich, 52425, J\"ulich, Germany}
\affiliation{MultiscaleConsulting, Wolfshovener str. 2, 52428 J\"ulich, Germany}
\author{F. Zhou}
\affiliation{State Key Laboratory of Solid Lubrication, Lanzhou Institute of Chemical Physics, Chinese Academy of Sciences, 730000 Lanzhou, China}
\author{B.N.J. Persson}
\affiliation{State Key Laboratory of Solid Lubrication, Lanzhou Institute of Chemical Physics, Chinese Academy of Sciences, 730000 Lanzhou, China}
\affiliation{Peter Gr\"unberg Institute (PGI-1), Forschungszentrum J\"ulich, 52425, J\"ulich, Germany}
\affiliation{MultiscaleConsulting, Wolfshovener str. 2, 52428 J\"ulich, Germany}

\begin{abstract}

We have studied the fluctuation (noise) in the position of sliding blocks under constant driving forces on different substrate surfaces. The experimental data are complemented by simulations using a simple spring-block model where the asperity contact regions are modeled by miniblocks connected to the big block by viscoelastic springs. The miniblocks experience forces that fluctuate randomly with the lateral position, simulating the interaction between asperities on the block and the substrate. The theoretical model provides displacement power spectra that agree well with the experimental results.

\end{abstract}

\maketitle

\thispagestyle{fancy}


{\it Introduction}--Studies of fluctuations in the position of particles, or in particle densities, are one of the most powerful methods to gain insight into the nature of particle systems. One of the first such studies was by Robert Brown (1827), who used a microscope to observe the motion of pollen grains in water. The phenomenon, later analyzed by Einstein \cite{Einstein}, provided convincing evidence for the existence of atoms and molecules. 

In this paper, we will investigate fluctuations in systems out of equilibrium, specifically in the position of a solid block sliding on a nominally flat surface under a constant driving force. We employ both experiments and simulations to examine this scenario. In this setup, the center of mass of the block moves with an average velocity $v$, but its actual position fluctuates around $vt$, i.e., the sliding distance is expressed as $s=vt+\xi(t)$, where $\xi(t)$ is a random variable with the time (or ensemble) average $\langle \xi \rangle = 0$, akin to the fluctuations observed in Brownian motion. From the obtained $\xi(t)$, we calculate the power spectra.

The surfaces of all solids exhibit surface roughness extending over many decades in length scale \cite{Cq,Nayak,earth0}. When two solids are brought into contact, they generally touch only a small fraction of the nominal contact area, whereas asperities make contact \cite{Ref1,Ref2,Ref3,Ref4,Ref5,Ref6,Ref7}. During sliding, asperity contact regions may undergo stick-slip motion, which may be correlated due to elastic coupling between different regions \cite{Ur,Nature,Mu1,Mu2,Fi}. The fluctuation $\xi(t)$ in the sliding distance arises from the stick-slip motion of the asperities, which is influenced by the nearly random nature of surface roughness.

Fluctuation (noise) in frictional sliding has been studied previously using two different methods. One method involves driving the slider at a constant speed and analyzing fluctuations in the driving force. However, accurately measuring these forces is challenging, making this method suitable only for systems with relatively large force fluctuations. In this study, we consider the case where the driving force is constant and study the fluctuations in the lateral position of the block. Distances can be measured accurately using various methods, with one extreme example being the laser displacement sensors used for studying gravitational waves, capable of measuring changes in distances down to $\sim 10^{-4}$ of the width of a proton \cite{gravitational}. A second method involves detecting and analyzing the sound waves emitted from the sliding junction \cite{sound,sound2}. However, correlating the sound wave frequency spectra to the motion of the block or the friction force acting on the block is not straightforward.

In this study, we use rubber blocks loaded with a constant normal force, sliding on different surfaces under a constant driving force. The substrates used are a (rough) concrete block and a (smooth) window glass surface. Since the glass surface is very smooth, a rubber asperity that contacts the glass surface at time $t=0$ will remain in contact with the glass for all times $t>0$. In contrast, the concrete surface, which is rougher than the rubber surface, causes the contact regions between the rubber and the concrete to change over time. Despite this qualitative difference, we find that for a given velocity $v$, the $\xi(t)$ power spectra are very similar but shifted along the frequency axis depending on the average sliding speed. 

This study may have implications for earthquake dynamics \cite{earthquake,earth1}. The phenomenological approach to earthquake dynamics uses rate-and-state friction models, which have their microscopic origin in the formation and breaking of asperity contact regions\cite{earth1,earth2}. The formation and breaking of asperity contact regions are also important in sliding electric contacts (e.g., in electric contact systems for trains) \cite{train,electric,electric1} and acoustic (acoustical noise) applications \cite{sound,sound2,Spurr}. Another related application is the dynamics of sand piles as a function of the tilt angle, which exhibits power-law power spectra \cite{sand}.

\begin{figure}[!ht]
\includegraphics[width=0.99\columnwidth]{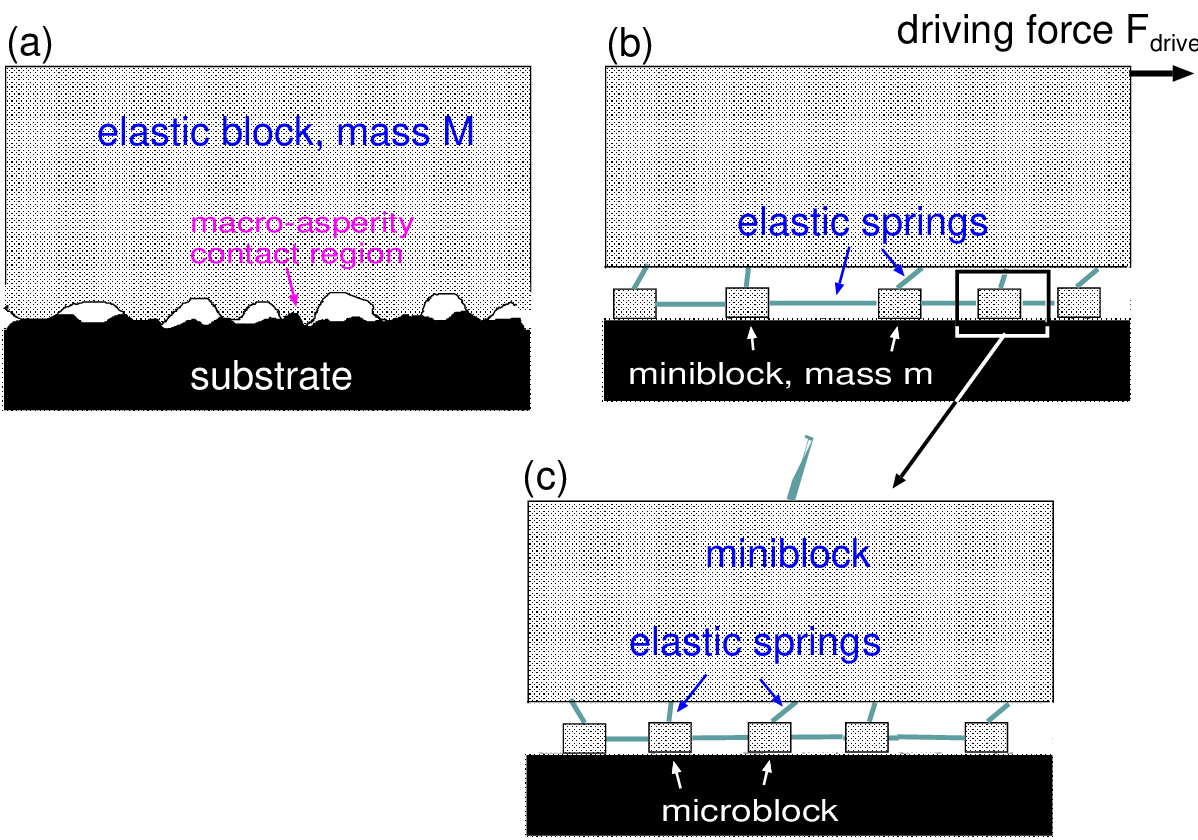}
\caption{\label{Model.eps}
A block-spring model. (a) The sliding block interacts with the substrate in $N$ asperity contact regions randomly distributed at the interface. The asperities experience randomly fluctuating forces from the substrate. (b) The model in (a) is replaced by a big block (mass $M$) connected to $N$ miniblocks with viscoelastic springs (spring constant $k_0$     and damping $\eta_0$). The miniblocks are coupled viscoelastically (spring constant $k_1$ and damping $\eta_1$) and experience randomly fluctuating (in time and space) forces from the substrate. (c) Real surfaces have roughness on many length scales, and the miniblocks in (b) are connected to smaller blocks (here denoted as microblocks), which will experience fluctuating forces from the interaction with the counter surface.}
\end{figure}

\begin{figure}[!ht]
\includegraphics[width=0.95\columnwidth]{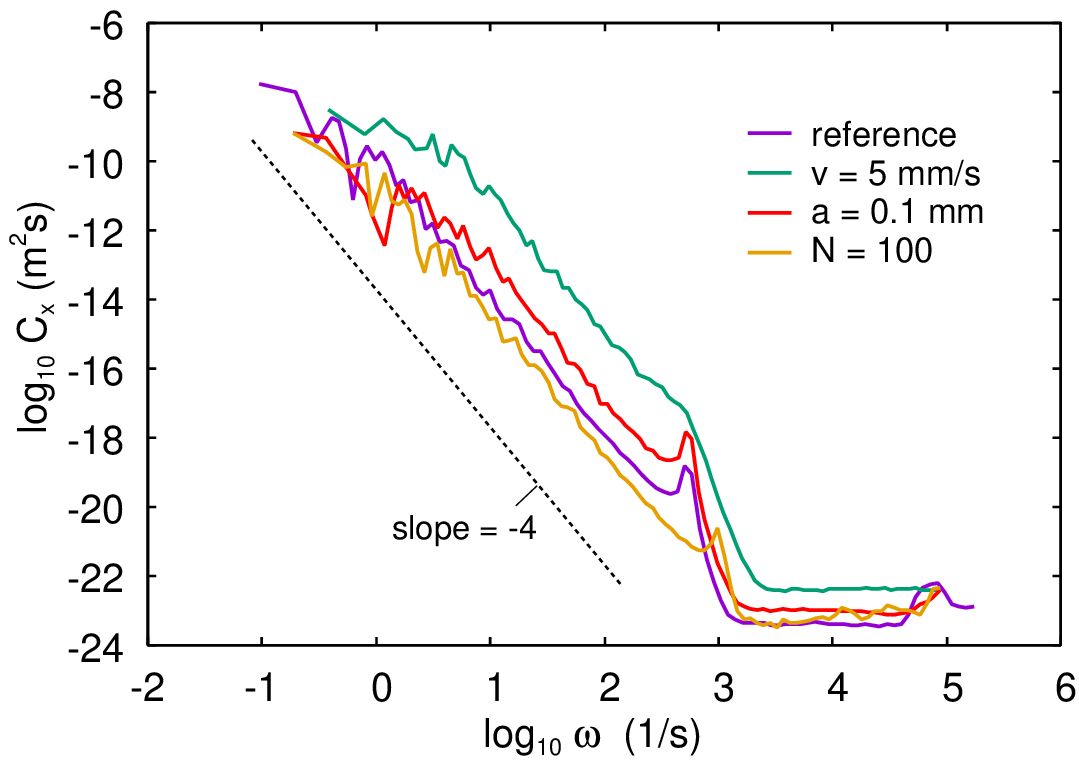}
\caption{\label{1logOmega.2logCx.theory.eps} 
The sliding distance power spectrum $C_x(\omega )$ as a function of the frequency. The reference case is for $N=30$ miniblocks and $v=0.5 \ {\rm mm/s}$ , and with $a= 1 \ {\rm mm}$, $F_{\rm drive}=10 \ {\rm N}$, $F_{\rm kin} = 5  \ {\rm N}$, $M = 1 \ {\rm kg}$, $m= 10^{-5}  \ {\rm kg}$, $k_0 = 10^4 \ {\rm Nm}$, $k_1 = 10^2 \ {\rm Nm}$ and $\eta_0 = 0.25 \surd (k_1/m_1)$. The other cases are the same as the reference case but with $v= 5 \ {\rm mm/s}$ (green curve), $a=0.1 \ {\rm mm}$ (red curve), and $N=100$ miniblocks (yellow curve). 
}
\end{figure}

\begin{figure}[!ht]
\includegraphics[width=0.95\columnwidth]{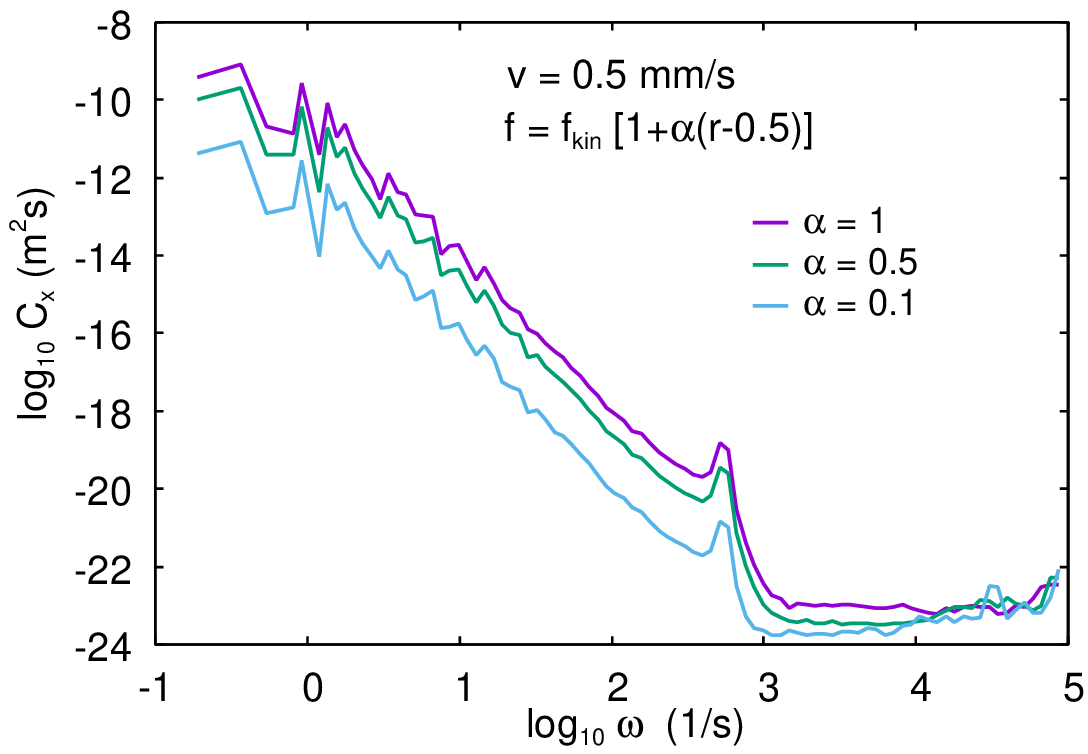}
\caption{\label{1logOmega.2logCx.vary.xf0.eps} 
The sliding distance power spectrum $C_x(\omega )$ as a function of the frequency. The results are for $30$ miniblocks for several values of the friction noise parameter $\alpha = 0.1$, $0.5$, and $1$. The other parameters are the same as for the reference case (see Fig. \ref{1logOmega.2logCx.theory.eps}).}
\end{figure}

\vskip 0.1cm
{\it Simulation model and results}--We consider the simple model shown in Fig. \ref{Model.eps}, which is similar to the Burridge and Knopoff model used in earthquake modeling \cite{earthquake,P}. Fig. \ref{Model.eps}(a) shows a block sliding on a randomly rough substrate, interacting with the substrate in $N$ asperity contact regions. The asperities experience randomly fluctuating forces from the substrate. We replace the model in (a) with the model in (b) where a big block (mass $M$) is connected to $N$ miniblocks with viscoelastic springs (spring constant $k_0$ and damping $\eta_0$). The miniblocks are coupled to nearby miniblocks viscoelastically (spring constant $k_1$ and damping $\eta_1$) and experience randomly fluctuating (in time and space) forces from the substrate.

Let $x(t)$ be the position of the large block and $x_i(t)$ the positions of the miniblocks ($i=1,...,N$). The velocity of the large block is $\dot{x} = dx/dt$. The equation of motion for the large block is
$$M \ddot{x} = F_{\rm drive} + F$$
where $F$ is the force from the miniblocks
$$F = -k_0 \sum_i (x - x_i) - m \eta_0 \sum_i (\dot{x} - \dot{x}_i)$$

The miniblocks obey the equations of motion
\begin{equation*}
    \begin{split}
        m\ddot{x}_i  = & k_1 (x_{i+1} + x_{i-1} - 2 x_i) + k_0 (x - x_i) \\
                       & - m \eta_1 \dot{x}_i - m \eta_0 (\dot{x}_i - \dot{x}) - f_{\rm kin} - f_i
    \end{split}
\end{equation*}

Here, $f_i(t)$ are randomly fluctuating forces with a time average $\langle f_i \rangle = 0$, and $f_{\rm kin}$ is the time-averaged kinetic friction force acting on each miniblock. We assume that $f_i(t)$ changes randomly with the sliding distance at an average rate denoted as $1/a$. If the large block moves from $x$ to $x+a$ during the time period $\Delta t$, the force on a miniblock (coordinate $x_i$) changes randomly between $t$ and $t+\Delta t$ from its old value to
$$f_i = \alpha f_{\rm kin} (r_i - 0.5)$$
where $r_i$ is a random number uniformly distributed between 0 and 1, and $\alpha$ is a parameter expected to be of order 1. These random fluctuations in $f_i$ are interpreted as resulting from changes in the contact between the asperities on the block and the substrate.
We define the position power spectrum as
$$C_x(\omega) = \frac{1}{2\pi} \int_{-\infty}^\infty dt \ \langle \xi(t) \xi(0) \rangle e^{i\omega t}$$

In the simulations presented below, we assume $F_{\rm drive} = 10 \ {\rm N}$ and $F_{\rm kin} = N f_{\rm kin} = 5 \ {\rm N}$, where $N$ is the total number of miniblocks. The spring constant $k_0 \approx ED$, where $D$ is the diameter of an asperity contact region. Using a rubber slider with $E \approx 10^{7} \ {\rm Pa}$ and assuming a typical diameter $D \approx 1 \ {\rm mm}$, we get $k_0 = 10^4 \ {\rm N/m}$. The mass of the large block is $M = 1 \ {\rm kg}$ and the mass of a miniblock is assumed to be a few times $\rho D^3 \approx 10^{-6} \ {\rm kg}$; we use $m = 10^{-5} \ {\rm kg}$. The lateral coupling between the miniblocks depends on the separation between macroasperity contact regions. Assuming a separation of order $D$, we get $k_1 \approx k_0$, but as the separation is likely larger, we take $k_1 = 10^2 \ {\rm N/m}$. However, simulations show that displacement power spectra are nearly the same for all $0 < k_1 < k_0$, so the exact value of $k_1$ is not critical for this study. The damping $\eta_0$ is chosen such that the vibrational motion of a contact region, if free, would be nearly overdamped, giving $\eta_0 \approx 
\surd (k_0/m)$; we use $\eta_0 = 0.8 \times 10^4 \ {\rm s}^{-1}$. The damping $\eta_1$ determines the increase in friction force with increasing sliding speed. We compare the theory to experimental data obtained for an average speed of $v = 0.5 \ {\rm mm/s}$, choosing $\eta_1$ so that the friction force equals $F_{\rm drive}$ at this speed. The sliding distance needed to renew the contact regions is set to $D$, so $a = 1 \ {\rm mm}$. The parameters above with $\alpha = 1$ and $N = 30$ are used as the ``standard" or ``reference" case. When parameters differ from this case, we will specify only the differing parameters.

The average frequency of fluctuations of the (random) force $f_i$ is $v/a$. For $a = 1 \ {\rm mm}$ and $v = 0.5 \ {\rm mm/s}$, this frequency is $0.5 \ {\rm s}^{-1}$. Another important frequency is the vibration frequency of the large block with miniblocks in fixed positions relative to the substrate, given by $\omega_{\rm c} = \surd (Nk_0 / M)$, which equals $548 \ {\rm s}^{-1}$ using our standard parameters ($N = 30$, $k_0 = 10^4 \ {\rm N/m}$, and $M = 1 \ {\rm kg}$). 
From the equation above one can show that for $v/a \ll \omega \ll \omega_{\rm c}$, we get \cite{notgiven}
$$C_x (\omega) \approx \frac{\alpha^2}{\omega^4} \frac{v^3}{12 \pi N a} \eqno(1)$$

Fig. \ref{1logOmega.2logCx.theory.eps} shows the sliding distance power spectrum $C_x(\omega)$ as a function of frequency. The violet curve represents $C_x(\omega)$ for the ``reference case" with $N=30$ miniblocks and $v=0.5 \ {\rm mm/s}$, using the parameters described above and in the figure caption. The other curves are for $v=5 \ {\rm mm/s}$ (green curve), $a=0.1 \ {\rm mm}$ (red curve), and $N=100$ miniblocks (yellow curve). Note that in all cases, there is a large frequency range where $C_x(\omega) \sim \omega^{-4}$.

Fig. \ref{1logOmega.2logCx.vary.xf0.eps} shows the sliding distance power spectrum $C_x(\omega)$ as a function of the frequency for several values of the friction noise-strength parameter $\alpha = 0.1$, $0.5$, and $1$. The other parameters are the same as for the reference case (see Fig. \ref{1logOmega.2logCx.theory.eps}).
The power spectra for $\omega < \omega_{\rm c}$ exhibit the scaling with $a$, $N$, $v$, and $\alpha$ as expected from the analytical result (1).

In Fig. \ref{1logOmega.2logCx.theory.eps} and \ref{1logOmega.2logCx.vary.xf0.eps}, the sharp peak at ${\rm log}_{10} \omega \approx 2.7$ or $\omega \approx 500 \ {\rm s}^{-1}$ is due to the big block vibrating against the substrate with the miniblocks at fixed positions. This resonance frequency is given by $\omega_{\rm c} = \surd (Nk_0 / M)$. Using $k_0 = 10^4 \ {\rm Nm}$, $M=1 \ {\rm kg}$, and $N=30$ gives $\omega_{\rm c} = 548 \ {\rm s}^{-1}$ in good agreement with the simulation results. For $N=100$, one gets $\omega_{\rm c} = 1000 \ {\rm s}^{-1}$ which also agrees with the simulations (see Fig. \ref{1logOmega.2logCx.theory.eps}). 

The broader peak starting at ${\rm log}_{10} \omega \approx 4.6$ or $\omega \approx 4 \times 10^4 \ {\rm s}^{-1}$ is due to a band of vibrations of the miniblocks against the substrate and the block. The resonance frequency of a single miniblock without friction is $\surd (k_0/m) \approx 3 \times 10^4 \ {\rm s}^{-1}$, but the miniblocks are coupled and experience friction towards the substrate and the block which influence and broaden the resonance. 

\begin{figure}[!ht]
\includegraphics[width=0.48\textwidth]{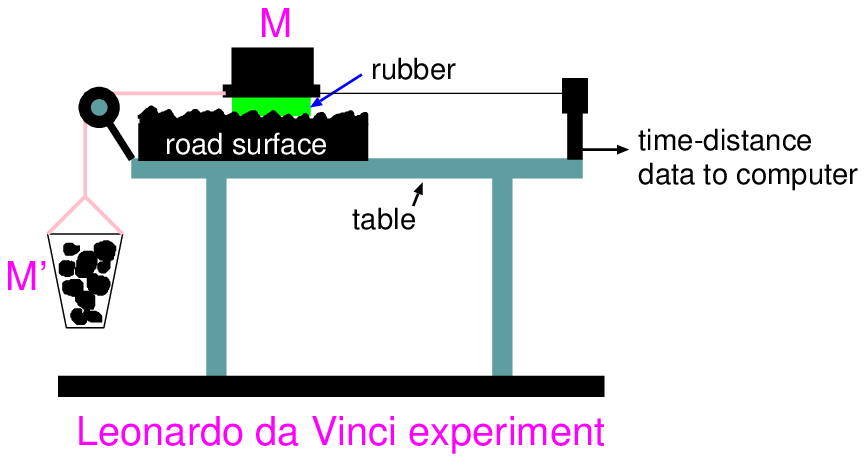}
\caption{\label{Leanardo.ps}
A simple friction slider (schematic) measures the sliding distance $x(t)$ via a displacement sensor.}
\end{figure}

\begin{figure}[!ht]
\includegraphics[width=0.95\columnwidth]{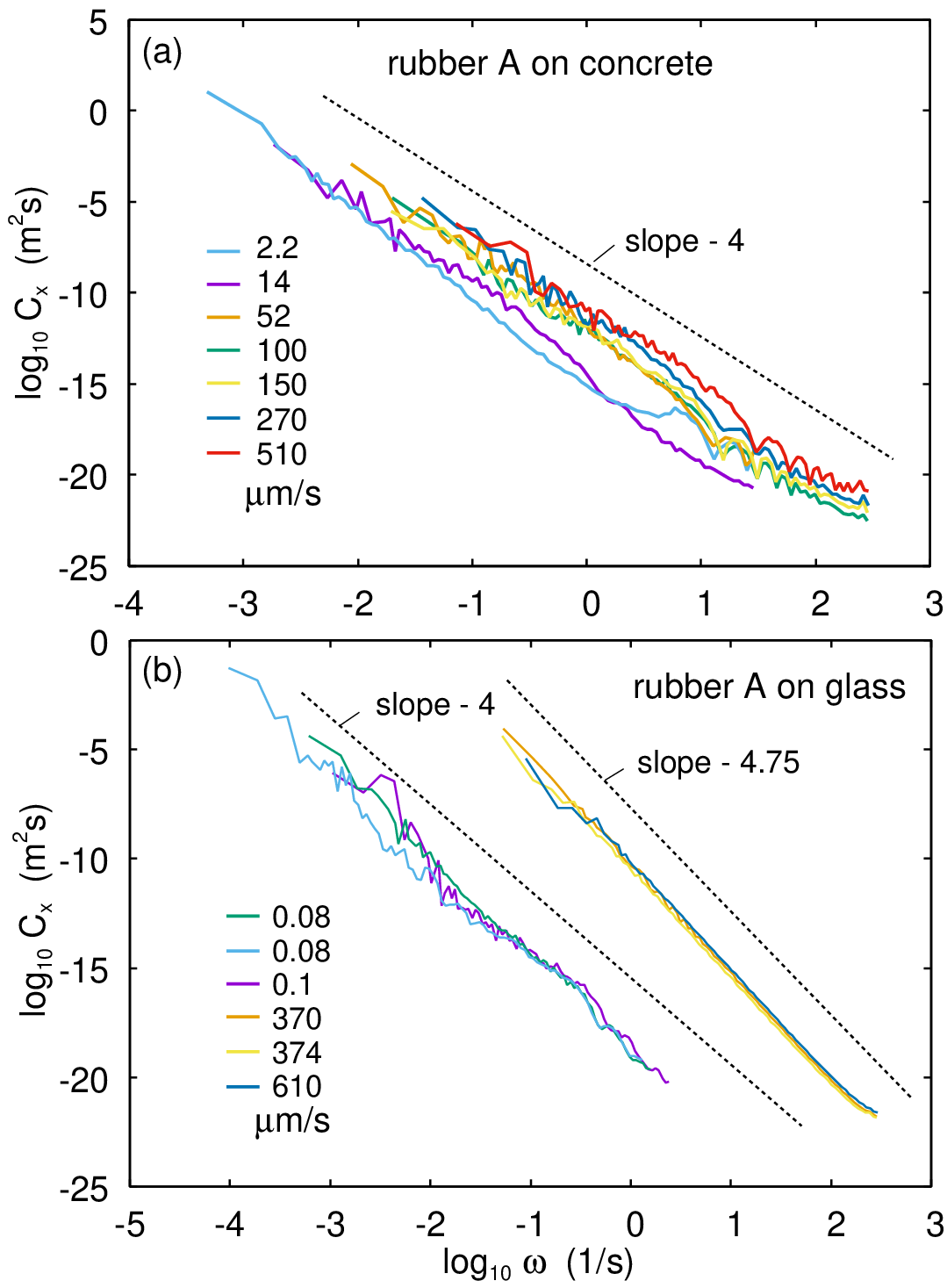}
\caption{\label{rubber_concrete.eps} 
The sliding distance power spectrum $C_x(\omega )$ as a function of the frequency for rubber blocks (compound A) sliding on (a) a rough concrete block and (b) a smooth silica glass plate, at different sliding speeds indicated.}
\end{figure}

\vskip 0.1cm
{\it Experiments and comparison with theory}--The experimental data were obtained using the setup shown in Fig. \ref{Leanardo.ps}. The slider consists of two rubber blocks glued to a wood plate, with one block positioned at the front and the other at the back. The nominal contact area is $A_0 \approx 10 \ {\rm cm}^2$. The normal force $F_{\rm N}$ is determined by the total mass $M$ of lead blocks placed on top of the wood plate. Similarly, the driving force is determined by the total mass $M'$ of lead blocks and the container (the mass of the ropes is neglected).

The sliding distance $x(t)$ as a function of time $t$ is measured using a Sony DK50NR5 displacement sensor with a resolution of $0.5 \ {\rm \mu m}$. This simple friction slider setup can also be used to calculate the friction coefficient $\mu = M'/M$ as a function of sliding velocity and nominal contact pressure $p = Mg/A_0$. 

We have measured the sliding distance $x(t)$ for rubber blocks sliding on a concrete block and on a (smooth) silica glass surface using different normal loads and driving forces. 

Fig. \ref{rubber_concrete.eps} shows the sliding distance power spectrum $C_x(\omega)$ as a function of frequency for a rubber block sliding on (a) the concrete block and (b) the silica glass plate at different sliding speeds, as indicated. Note that the slope of the curves for the concrete surface is about $-4$, so the low-frequency power spectra are approximately proportional to $\omega^{-4}$. For the glass surface at high sliding speeds, the power spectrum displacement exponent is approximately $-4.75$, while at low sliding speeds, it is the same as for the concrete surface. Additional measurements on the smooth glass surface using another rubber compound (compound B, result not shown) revealed a displacement exponent of approximately $-5$. This indicates that different interfacial processes may occur on the glass surface compared to the concrete surface, possibly related to the influence of contamination or rubber wear particles on the sliding motion.

\begin{figure}[!ht]
\includegraphics[width=0.95\columnwidth]{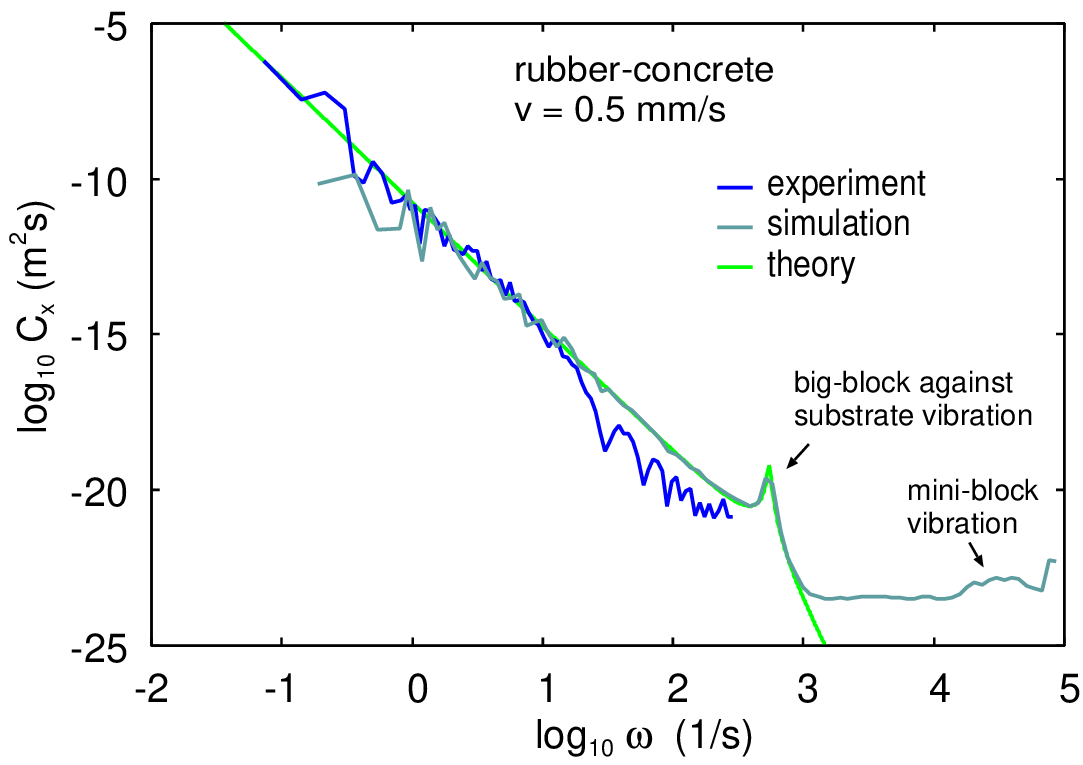}
\caption{\label{1logOmega.2logCx.theory.experiment.concrete.eps} 
The sliding distance power spectrum $C_x(\omega )$ as a function of the frequency. The experimental result is for a rubber block sliding on a concrete surface, the theory result is for the reference case with $N=30$ miniblocks and for $v=0.5 \ {\rm mm/s}$, $a=1 \ {\rm mm}$ and $\alpha=0.4$.
}
\end{figure}

Fig. \ref{1logOmega.2logCx.theory.experiment.concrete.eps} shows the sliding distance power spectrum $C_x(\omega)$ as a function of frequency for a sliding speed of $v=0.5 \ {\rm mm/s}$. The experimental result is for a rubber block sliding on a concrete surface. The theoretical results (green and gray curves) are for the reference case with $N=30$ miniblocks, $v=0.5 \ {\rm mm/s}$, $a=1 \ {\rm mm}$, and $\alpha=0.4$. Note that the experimental data exhibit the same $\sim \omega^{-4}$ scaling as the theoretical curve.

In the simulation, a high-frequency roll-off region occurs due to the damped motion of the miniblocks. No such region is observed in the experiments, but this may be due to the limit in the frequency region we can probe with the present experimental setup.

In an earlier study, the fluctuation in the friction force for metal-metal contacts was measured using a pin-on-disc setup (pin from aluminum and steel disc) \cite{Louis}. The sliding speed was constant at $v=1 \ {\rm cm/s}$, and the friction force was studied as a function of time. The force power spectra exhibited approximately a $\sim \omega^{-1}$ frequency dependency in a low-frequency interval, differing from our result for concrete where the position coordinates fluctuate with a $\sim \omega^{-4}$ power spectrum and the force (which is proportional to $\omega^4 C_x(\omega)$) with a $\sim \omega^0$ power spectrum. However, the authors in Ref. \cite{Louis} attributed the friction force fluctuations to wear particles. When the wear particles were removed, the force power spectrum became close to $\sim \omega^0$, consistent with our findings. Additional tests using another rubber compound (compound B, result not shown) showed a $\sim \omega^{-5}$ displacement power spectrum and a $\sim \omega^{-1}$ force power spectrum which we believe is due to surface contamination or wear, also aligning with the conclusions of the study in Ref. \cite{Louis}.

\vspace{0.1cm}
{\it Conclusions and outlook}--During the sliding of a solid block on a substrate, asperity contact regions form and disappear in a nearly random manner. Even if the time-averaged friction force remains constant during constant speed sliding, the friction force will exhibit fluctuations around its time-averaged value. Consequently, if the applied driving force is constant, the sliding velocity will fluctuate around its time-averaged value. We have experimentally studied these fluctuations for rubber blocks sliding on concrete and glass surfaces. For sliding on concrete the power spectrum of the fluctuations in the position of the sliding block exhibits a $\omega^{-4}$ frequency dependency. We can reproduce this $\omega^{-4}$ frequency dependency using a simple block-spring model with fluctuating forces associated with the breaking and formation of asperity contact regions. 
For sliding on smooth glass surfaces the displacement exponent is either -4 or close to -5 depending on the rubber compound and the experimental conditions; we associate these fluctuations in the displacement exponent to 
wear particles and contamination films at the sliding interfaces.

We plan to measure the sliding distance with a displacement sensor that has a much higher resolution. This would allow us to probe the block position at much higher frequencies than is possible with our current instrument. With such a setup, it could also be possible to obtain information about the onset of sliding, involving the transition from static friction to kinetic friction, which is of great interest for earthquake dynamics.
To model this, it may be necessary to extend the model to include the hierarchical nature of real surface roughness, with smaller asperities on top of larger asperities. We plan to study this using a hierarchical distribution of sliding blocks, with smaller blocks attached to larger blocks, and so on (see Fig. \ref{Model.eps}).


\begin{thebibliography}{99}
\bibitem{notgiven}
To be published by the authors.
\end{thebibliography}
\end{document}